\documentclass[12pt,preprint]{aastex}
\usepackage{amsmath}
\usepackage{psfig}

\eqsecnum

\renewcommand\email\texttt

\def\spose#1{\hbox to 0pt{#1\hss}}
\def\lta{\mathrel{\spose{\lower 3pt\hbox{$\sim$}}
    \raise 2.0pt\hbox{$<$}}}
\def\gta{\mathrel{\spose{\lower 3pt\hbox{$\sim$}}
    \raise 2.0pt\hbox{$>$}}}

\font\sc=cmcsc10
\def\Msun{\,M_\odot}
\def\Lsun{{\rm L}_{\rm \odot}}

\input epsf
\begin{document} 

\shorttitle{\sc Observed properties of dark matter}
\shortauthors{Gilmore etal}

\title{The Observed properties of Dark Matter on small spatial scales}

\author
{Gerard Gilmore\altaffilmark{1}, Mark I. Wilkinson\altaffilmark{1,6},
 Rosemary F.G. Wyse\altaffilmark{2}, Jan T. Kleyna\altaffilmark{3},
 Andreas Koch\altaffilmark{4,5}, N. Wyn Evans\altaffilmark{1}, Eva
 K.~Grebel\altaffilmark{5,7} } 
\altaffiltext{1}{Institute of Astronomy, University of Cambridge,
Madingley Road, Cambridge CB3 0HA, UK;\email{gil@ast.cam.ac.uk}}
\altaffiltext{2}{The Johns Hopkins University, Department of Physics
 \& Astronomy, 3900 N.~Charles Street,  Baltimore, MD 21218, USA}
\altaffiltext{3}{Institute for Astronomy, University of Hawaii, 2680
  Woodlawn Drive, Honolulu, HI 96822, USA}
\altaffiltext{4}{Department of Physics and Astronomy, UCLA, 430
 Portola Plaza, Los Angeles, CA 90095-1547, USA}
\altaffiltext{5}{Astronomical Insitute of the University of Basel, 
Department of Physics and Astronomy, Venusstr. 7, CH-4102 Binningen, Switzerland}
\altaffiltext{6}{Department of Physics and Astronomy, University of
  Leicester, University Road, Leicester, LE1 7RH, UK.} 
\altaffiltext{7}{Astronomisches Rechen-Institut, Zentrum f\"ur Astronomie der
Universit\"at Heidelberg, M\"onchhofstr.\ 12-14, D-69120 Heidelberg,
Germany.}

\begin{abstract} 
There has long been evidence that low-mass galaxies are systematically
larger in radius, of lower central stellar mass density, and of lower central
phase-space density, than are star clusters of the same luminosity.
The larger radius, at a comparable value of central velocity
dispersion, implies a larger mass at similar luminosity, and hence
significant dark matter, in dwarf galaxies, compared to no dark matter
in star clusters.  We present a synthesis of recent photometric and
kinematic data for several of the most dark-matter dominated
galaxies. There is a bimodal distribution in half-light radii, with
stable star clusters always being smaller than $\sim30$pc, while stable
galaxies are always larger than $\sim120$pc.  We extend the
previously known observational relationships and interpret them in
terms of a more fundamental pair of intrinsic properties of dark
matter itself: dark matter forms cored mass distributions, with a core
scale length of greater than about 100pc, and always has a maximum central mass
density with a narrow range. The dark matter in dSph galaxies appears to be
clustered such that there is a mean volume mass density within the
stellar distribution which has the very low value of about 0.1$\Msun$
pc$^{-3}$ (about 5GeV/c$^2$ cm$^{-3}$). All dSphs have velocity
dispersions equivalent to circular velocities at the edge of their
light distributions of $\sim 15$km\,s$^{-1}$. In two dSphs
there is evidence that the density profile is shallow (cored) in the
inner regions, and so far none of the dSphs display kinematics which
require the presence of an inner cusp. The maximum central dark matter
density derived is model dependent, but is likely to have a mean value
(averaged over a volume of radius $10$pc) of $\sim0.1\Msun$ pc$^{-3}$
(about $5$GeV/c$^2$ cm$^{-3}$) for our proposed cored dark mass
distributions (where it is similar to the mean value), or
$\sim60\Msun$ pc$^{-3}$ (about $2$TeV/c$^2$ cm$^{-3}$) if the dark
matter density distribution is cusped. Galaxies are embedded in dark
matter halos with these properties; smaller systems containing dark
matter are not observed. These  values provide new information
into the nature of the dominant form of dark matter.
\end{abstract}

\keywords{
dark matter---galaxies: individual (dSph)---galaxies:
kinematics and dynamics---Local Group---stellar dynamics}

\section{Introduction}

The distributions of total luminosity and of central stellar velocity
dispersion for star clusters and for dwarf galaxies overlap, so that
the faintest galaxies have approximately the same values of these
physical parameters as do star clusters, with galaxy luminosities
extending as faint as $\sim 10^3$L$_\odot$, with line-of-sight central
velocity dispersions of $\sim 10$~km/s.  The half-light radii (radius
containing one-half the total luminosity) of the galaxies, however,
are significantly larger (hundreds of parsec) than those of star
clusters (at most tens of parsec). This leads, through the virial
theorem, to significantly larger inferred masses for the dwarf
galaxies, compared to star clusters of the same luminosity and
velocity dispersion. Indeed, the derived values of central and global
mass-to-light ratios for the gas-poor, low-luminosity, low-surface
brightness satellite galaxies (classified as Dwarf Spheroidal
galaxies; dSph) of the Milky Way are high, up to several hundred in
solar units, making these systems the most dark-matter-dominated
galaxies in the local Universe (e.g.~Mateo 1998 for a convenient
review of early work).  As we discuss further below, they are the ideal
test-beds for constraining the nature of the dark matter that
dominates their gravity \citep{jpops03}.

The dSph galaxies and star clusters share a further observed property,
that organised orbital rotational energy of the member stars is
negligible compared to the energy in disordered motion, which is
measured by the stellar velocity dispersion at a given location.
Similarly to pressure gradients in a fluid, the stellar velocity
dispersion provides the support against self-gravity, but unlike the
fluid case, stellar pressure can be anisotropic, generating galaxy
shapes which need not be spherical.  Systems in which angular momentum
support against gravitational potential gradients can be ignored when
analysing the kinematics of member stars are designated as `hot'.

It has been known for the past twenty years that there are
well-defined, and probably fundamental, scaling relations between the
half-light radius (or core radius), the central velocity dispersion
and the luminosity of hot stellar systems (e.g. Kormendy 1985; Bender,
Burstein \& Faber 1992; Zaritsky, Gonzalez \& Zabludoff 2006a,b). It
has further been long-established that the globular star clusters in
the halo of our Galaxy show distinctly different scalings from the
dSph galaxies, and that the dSph galaxies in turn have different
scalings from more luminous hot galaxies (e.g. Kormendy 1985, his
Figure 3; Burstein et al.~1997).  Dynamical effects over their long
lives have modified the size and luminosity distributions of the
Galactic globular clusters (e.g.~Fall \& Rees 1977; Gnedin \& Ostriker
1997), so it is important that robust studies include star clusters of
all ages and in all environments, including globular star clusters in
external galaxies, nuclear star clusters, and young massive star
clusters, significantly younger than globular clusters (e.g.~Walcher
et al.~2005; Seth et al.~2006).  

The specific combination of central velocity dispersion ($\sigma_0)$
and half light radius ($r_h$)  $r_h^{-2}\sigma_0^{-1} \propto \rho_h
\sigma_0^{-3}$, is a convenient measure of the phase-space density,
where $\rho_h$ is the mean density within a half-light radius.  The
similarity of velocity dispersion for star clusters and dwarf galaxies
of the same luminosity, combined with the factor of $\sim $~ten
difference in their half-light radii, implies a systematic difference of
some two orders of magnitude in the value of the phase-space density
at fixed stellar mass (see e.g.~Walcher et al.~2006).

What is the physical explanation for these differences between star
clusters and low-luminosity galaxies? Clearly, the presence of dark
matter in dSph galaxies, but not in star clusters, is a critical
distinction, and provides an opportunity to identify underlying
physics of Dark Matter. For example, the suggestion by Mateo et
al.~(1993) that there is an apparent minimum dark halo mass of $\sim
2\times10^7\Msun$, deduced from the available dynamical studies,
implies a small-scale limit to the dark matter power spectrum unlike
that assumed in $\Lambda$CDM models (e.g.~Moore et al.~1999; Klypin et
al.~1999). The minimum halo mass suggestion was shown to still be
valid in a significantly extended sample, including dSph satellites of
M31, in addition to those of the Milky Way, in an important study by
C\^ot\'e et al.~(1999; see especially their figure~3). C\^ot\'e et
al.~also provided one of the earliest robust demonstrations that the
internal kinematics of dSph galaxies are in general unaffected by
external tidal forces from their host galaxies, so that the results
from application of equilibrium dynamical analyses are reliable.  The
putative minimum dark-halo mass was still found to be appropriate, in
the larger sample with more extensive data reviewed by \cite{W06a}
and by \cite{LADM}.

Within the broad class of dwarf galaxies, of which dSph are the
least luminous members, one can apply simple models to the data and
obtain scaling relations between such quantities as derived central
dark matter density and observed central velocity dispersion
(e.g.~Kormendy \& Freeman 2004).  The dSph galaxies are systematically
discrepant in their correlation fit, falling below the extrapolated
trend to larger central densities as luminosities decrease. We provide
an explanation here by showing that the dSph form the limit of such
relations, not a continuation.  These correlations can be used to
consider compatability with various parameterizations of the power
spectrum of primordial density fluctuations.  Dwarf galaxies play a
special role, in that they appear to be the smallest systems in which
dark matter dominates, and so provide a powerful test of the power
spectrum on the smallest scales. The smallest scales on which dark
matter particles cluster depends on the physical characteristics of
the dark matter itself (e.g. Green, Hofman \& Schwarz 2005).
Determining this smallest scale is the goal of the present analysis.

In this paper we revisit the established correlations and scaling
relations for dwarf galaxies and for star clusters. Stellar velocity
data now exist for stars across the face of several of the dSph,
allowing an analysis that goes well beyond that possible with just the
central value of the velocity dispersion, a limitation in early
studies. The discussion below takes account of these new data, where
available. Recent imaging data allow a re-evaluation of the sizes of
star clusters and galaxies, strengthening the case for a real
discontinuity between star clusters and galaxies.  We interpret our
findings in terms of a more fundamental pair of intrinsic properties
of dark matter itself.

\section{The sizes and internal kinematics of star clusters and galaxies}

The existence of a clear observational distinction between massive
star clusters and low-mass galaxies has been substantially
strengthened recently, both through detailed studies of more luminous
and massive star clusters in a wide range of environments, and through
discovery of a large number of extremely low-luminosity satellite
galaxies around the Milky Way (e.g.~Willman et al.~2005b; Belokurov et
al.~2007; Zucker et al 2006a, 2006c) and around M31 (e.g.~Zucker et al
2004, 2006b), mostly based on imaging from the Sloan
Digital Sky Survey (SDSS). Fig.~1 shows the current sample in a plot
of half-light radius against absolute magnitude in the V-band.  Star
clusters in all studied environments, with luminosities over the whole
range from M$_V = -4$, L$\sim 10^3$~L$_\odot$, up to M$_V=-15$,
L$\sim10^9\, \Lsun$, and a wide range of ages, invariably have
characteristic scale sizes $r_{h}$ less than about 30pc.  Thus the
dynamical range over which both star clusters and galaxies exist, and
over which there is a distinct size dichotomy, has now been
established to cover some six orders of magnitude in stellar
luminosity. Available direct studies of the stellar initial mass
function in star clusters and dSph galaxies (e.g.~Wyse et al.~2002) show
this range in luminosity corresponds to a similar dynamic range in
baryonic mass.

Figure~1 makes evident that there is a robust maximum radius to star
clusters, at all luminosities.  Figure~1 also illustrates that the
dSph galaxies in both the Milky Way and in M31 have a minimum
characteristic radius, and this minimum is a factor of $\gta4$ larger
than the largest star clusters.  With the exception of the recently
discovered object ComaBer (Belokurov et al.~2007) -- the largest and
brightest of the three systems indicated by ringed red circles in
Fig.~1 -- which manifestly merits further study, but is one of the two
new dSph which lie between the Sgr dwarf and the Magellanic Clouds,
and which show significant indications of tidal disruption, there is no known
object in the size-gap between $\sim30$pc and $\sim120$pc.

It is more correct to say that, modulo ComBer, there is no known
stable object in the size gap. This intermediate size might be
occupied transiently by a larger object (a dwarf galaxy) in the very
late stages of disruption by external (Galactic) tides, or by a small
object (globular cluster) in the last stages of evaporation. In the
first of these cases a low velocity dispersion compact core can be
generated transiently, if outer, hotter, stars are removed by a
suitable tide, while in the second case the density profile changes
systematically from the small value typical of a compact star cluster
to a a very large value, almost constant density, covering all
possible radii during that (short-lived) process.

Willman~1~\citep[radius 20pc;][]{Wi06} and
Segue~1~\citep[radius 30pc;][]{catsdogs} are also newly discovered and
interesting tests of the conclusions of this section, with Segue~1
showing evidence for significant tidal disruption. The two largest
Galactic globular clusters are Pal 14, with size 28~pc, and Pal 5,
24~pc, which is in an advanced stage of tidal disruption with
prominent streams of stripped stars (\cite{Od2003}). The largest Ultra
Compact Dwarf galaxies (UCDs),
also with size 25pc, are associated with the centre of the Virgo
cluster, and the galaxy M87, and were for a long time suspected
\citep{Dr03, Ha05} of being small galaxies severely affected by
tides. Our interpretation of their status, based on their position in
Figure~1, is that they are simply very massive star clusters, with no
associated dark matter. 

There are two very recent detailed dynamical analyses of the masses of
ultra compact dwarf galaxies, by \citet{Hilker07} and by
\citet{EGDH07}. \citet{Hilker07} derived dynamical masses for five
ultra-compact dwarfs and one dE nucleus in Fornax, while
\citet{EGDH07} studied six Virgo UCDs and five very luminous Fornax
UCDs. They show that all these systems are similar, in structure and
dynamics, and that the dynamical mass-to-light ratios for the UCDs are
consistent with simple stellar models: there is no evidence for any
dark matter associated with these stellar clusters.  They are the
(very) high-mass/high-luminosity extreme of more typical globular
cluster populations.

There are systematic effects that need to be taken into consideration
when interpreting Figure~1.  First, the half-light radius (whose
definition is the obvious one, the radius enclosing one-half the total
luminosity) can be robustly estimated only in systems with a
well-defined and convergent luminosity profile. In many cases the
parameter published is a `core radius', that radius at which the
projected surface brightness has fallen to one-half its central
value. For the commonly used King-model fits to star clusters, the
core and half-light radii are similar, except in (low concentration)
cases where the object has an extended tail to the brightness
distribution. In this case, the derived core radius underestimates the
true half-light radius.  For a modelled projected surface brightness
$\mu(r)$ described by 
\begin{equation} 
\mu(r) = \mu_0\left(1 + r^2/a^2\right)^{-\gamma/2},
\end{equation}
where $a$ is the scalelength of the core, $\mu_0$ the
central surface density and $\gamma$ the power-law decline of the
surface density at large radii, King models have $\gamma \sim2$, and
the Plummer sphere has $\gamma =4$. The projected integrated
luminosity $L_p$ is 
\begin{equation}
L_p(r) = \frac{2\pi \mu_0}{\gamma - 2}\left(a^2 - a^{\gamma}\left(a^2
  + r^2\right)^{-\frac{(\gamma - 2)}{2}}\right),
\end{equation} 
where the integral to infinity is
\begin {equation}
L_{tot} = \frac{2 \pi \mu_0 a^2}{\gamma - 2}.
\end{equation} 
For the Plummer sphere $L_p(a)=1/2L_{tot}$. 
That is, the physical meaning of the scale parameter
$a$ in this Plummer case is a half-light radius.  More generally, 
the relation between the scale parameter $a$ and the
core radius $r_c$, defined as above, can be shown to be
\begin{equation}
r_c = a(2^{2/\gamma} - 1)^{1/2},
\end{equation}
where $a$ is the (cylindrical) radius which encloses one-half the
total luminosity, so that for a Plummer sphere observed in projection
the half-light radius $a$ and core radius $r_c$ are related by $a
\approx 3r_c/2$. Core radii, where fitted, are adequate approximations
to half-light radii for King model globular star clusters, and are
lower limits to half-light radii for dSph galaxies, while the Plummer
scale parameter is a half-light radius, to the accuracy of a Plummer
model fit to the data.  Conservatively, we adopt and identify three
cases of core radii for dSph galaxies at face value, rather than
converting to larger, but more uncertain, half-light radii.

Second, only the extent of the baryonic component is being measured,
and there is no guarantee -- or need -- for mass to follow light.
Given that most, if not all (as we argue here), galaxies are embedded
in extended dark matter halos, photometric determinations are probably
a lower limit on the scale length of the total mass distribution (we
discuss this below for two specific caes, UMi and Fornax). Figure~1 is
therefore probably conservative in describing mass: an even larger
distinction between star clusters and dSph galaxies would be seen were
one able to plot parameters describing mass rather than light.

Figure~1 is further conservative, in that it is possible that the
recently discovered very-low luminosity dSph are in fact larger than
is shown.  The recent history of observational studies of nearby
low-luminosity galaxies, in which individual stars are resolved and
the extent on the sky is measured through star counts, corrected for
foreground stars in the Milky Way Galaxy, has been that their radial
extents (and hence total luminosities and half-light radii) tend to be
somewhat under-estimated (eg \cite{Od2001}). As photometric data are
extended to lower surface brightnesses, and as kinematic studies of
individual member stars develop, allowing foreground stars to be
rejected on the basis of line-of-sight velocity, the galaxies are
typically found to be larger than first measured.  For example, for
the best-studied dSph the estimated total extent has changed by a
factor 1.5--2 over the last 10 years. In contrast, star clusters
really do have steep outer light profiles, and thus have well-defined
observational parameters.  Both these effects lead to a systematic
observational underestimate of any gap in spatial scale between star
clusters and galaxies.

\section{Masses and mass distributions, cores and cusps}

It has long been known that the observed value of $\sim 10$~km/s for
the line-of-sight velocity dispersions of Local Group dwarf
spheroidals (dSphs), together with their $\sim200$pc half-light radii,
implies mass-to-light ratios $M/L$ of up to $\gta 100 \Msun/\Lsun$.
Until recently, most of these estimates of $M/L$ were based upon a
measurement of only the central value of the velocity dispersion, and
upon the assumption that the mass profile follows the light profile.
The availability of datasets of radial velocities for hundreds of
individual stars spread out in radius across the nearby dSphs,
obtained by several groups, has changed all this.  To date, the
line-of-sight velocity dispersion profiles in the Fornax, Draco, Ursa
Minor, Carina, Leo~I, Leo~II, Sextans and Sculptor dSphs have been
mapped to the (rather poorly defined concept of an) optical
edge~\citep{mateo97, kleyna02, W04, W06a, Munoz2005, Sohn2006, Koch06,
Batt06, Walker06}  - see Figure~2 for a sample of profiles obtained by our
group.   Central velocity dispersions have been derived for
several of the newly discovered extremely low-luminosity dSph
satellites of both the Milky Way and M31 [\cite{KWEG05, ICILM06,
Munoz06b}]. Several major studies are underway so that all the known dSph
galaxy satellites with $ -13 \lta {\rm M_V} \lta -8$ will soon
have much improved determinations of their dynamical mass
distributions inside their optical radii, while some information will
be available on the several dSph galaxies with  $ -8 \lta {\rm M_V}
\lta -4$. Our conclusions in this paper lead us to predict extended
dark matter halos for the systems with characteristic radii that place
them in the `galaxy' regime of Fig.~1.  Future more extensive studies
of these very faint galaxies, as well as continuing discoveries, will
allow our predictions to be tested.

\subsection{Mass modelling for hot stellar systems}

Stellar systems with no net internal angular momentum maintain their
scale size by the pressure support of random stellar motions against
the gravitational potential gradient. This pressure is naturally
tri-axial in a collisionless system. A robust dynamical analysis of
such a system involves solution of the Collisionless Boltzmann
Equation (CBE) for some appropriate (stellar) tracer particle
phase-space distribution function, determining the mass distribution
which generates the gravitational potential gradients along which the
stars orbit. Such models are being applied and further developed
[\cite{kleyna01, wilkinson02, Walker06}] to the few best-studied dSph
galaxies. Such analyses are appropriate for, and require, velocity and
position data for (at least) several hundred tracer stars distributed
across the dSph. For most dSph galaxies studied to date, the more
limited data available justifies analysis only of the first velocity
moment of the distribution function, the velocity dispersion as a
function of radius.

In a collisionless equilibrium system the Jeans' equations are the
relation between the kinematics of the tracer stellar population and
the underlying (stellar plus dark) mass distribution. In terms of the
intrinsic quantities, and assuming spherical symmetry, the mass
profile may be derived as:
\begin{equation*}
M(r) = -\frac{r^2}{G}\left( \frac{1}{\nu}\frac{{\rm d}\,
\nu\sigma_r^2}{{\rm d}\,r} + 2\, \frac{\beta\sigma_r^2}{r} \right) 
\end{equation*}
where $\sigma_r(r)$ is the one-dimensional stellar velocity-dispersion
component radially towards the center of the mass distribution,
$\beta(r)$ quantifies the stress term associated with (possibly
radially variable) velocity orbital anisotropy, and $\nu(r)$ is the
stellar density distribution.

The quantities directly observed are the line-of-sight velocity
dispersion as a function of projected radius, $R$, $\sigma(R) =
<v(R)^2>^{1/2}$ and the surface brightness profile as a function of
projected radius.  Given finite amounts of data defining the projected
surface brightness and kinematic distribution functions, we can
proceed in either of two ways.  We may assume {\sl a priori\/} a
parameterised mass model $M(r)$ and velocity anisotropy $\beta(r)$ and
fit the observed velocity dispersion profile; or we may use the Jeans'
equations to determine the mass profile from the projected velocity
dispersion profile, utilising some (differentiable) functional fit to
the observed light distribution and a (range of) assumed form(s) for
the anisotropy $\beta(r)$. Assuming spherical symmetry, it is
straightforward to obtain $\langle \sigma_r^2 \rangle$ from the
observed line-of-sight velocity dispersion using Abel integrals. In
what follows, we take the second approach to the Jeans' equation
analysis: both the spatially binned dispersion profile and the surface
brightness distribution are fit by an appropriate smooth function, and
we assume an isotropic velocity dispersion. Figure~\ref{fig:JeansFits}
shows some examples of the fits to the light and dispersion profiles
used in the analysis.

 It is obvious from the Jeans' equation that radially variable
velocity dispersion anisotropy is degenerate with mass, making any
deductions as to whether or not the inner mass profile is cored or
cusped in general model-dependent. Further information is needed to
break this degeneracy, and fortunately is sometimes available, as we
discuss below. In general however, full multi-component
distribution-function models using adequately large datasets, as
discussed in Section 3.4 below, are required to use the information in
the data to break this degeneracy.

\subsection{Moment equation analyses of inner dark mass distributions}

Jeans' equation dynamical analyses generate three quantities. The most
robust is the {\it mean\/} dark matter mass density inside the radius
where adequate kinematic data are available. Similarly robust is the
total mass, again inside the radius where adequate kinematic data are
available. The analysis can also constrain the mass density in a small
central region, though the limited central spatial sampling makes this
less robust, and dependent on an adopted underlying mass density
profile.  It remains the case, of course, that a suitably small
central cusp in the mass distribution would not be detected so long as
it was unresolved inside the available kinematic data: very small cusps
might be present. 

Our Jeans' analysis mass models are presented in Figure~4. Some
caution is required in the interpretation of these profiles, given the
simplifying assumptions which have been made (spherical symmetry;
velocity isotropy; smooth dispersion and light
profiles). First, given that satisfying the Jeans equations is a
necessary, but not sufficient, condition for a solution of the CBE to
be everywhere non-negative (and hence a viable distribution function),
the models presented here are not guaranteed to correspond to physical
models. However, we note that for a tracer distribution with an
isotropic velocity distribution and density profile $\nu \sim
r^{-\gamma}$, the logarithmic slope of any external power-law
potential $\psi \sim r^{-\delta}$ must satisfy $\delta\leq 2\gamma$ to
ensure the non-negativity of the distribution
function~\citep{AnEvans06}. Thus, it is reasonable to assume that for
a tracer distribution with $\gamma \sim 0$, a cored mass density
distribution will yield a physically meaningful distribution
function. Second, in this analysis we have assumed specific forms for
the light distribution and the mass profiles obtained will be
sensitive to these assumed forms. However, as
Figure~\ref{fig:JeansFits} illustrates, our profiles constitute
reasonable representations of the observed data - but it is worth recalling
that the very innermost light profiles of the dSphs are often poorly
defined. 

Our representation of the velocity dispersion profiles as smooth
functions which are flat in the innermost regions may mask
features which are visible in the profiles at marginal
significance. It has been known for many years that the stellar
populations in dSph are complex, with the implications of this
complexity evident in many analyses, yet difficult to model fully
without very large data sets. With respect to a superposition of two
populations, one should be careful when comparing this to the observed
star formation histories.  Episodic star formation with clearly
distinguished periods of star formation has been detected only in
Carina; all other dSphs with extended star formation histories show
evidence of long-lasting activity without obvious pauses.  So one is
not really dealing with, e.g., only two distinct populations (as is
sometimes claimed in the literature).  It can be shown, however, that
younger and/or more metal-rich populations are more centrally
concentrated (e.g., Harbeck et al.  (2001)), and that these
populations also tend to exhibit lower velocity dispersions (Tolstoy
et al. (2004)).  Recently, \cite{McConnachie07} have recalled the
general point that a rising dispersion profile (e.g. Leo~I) might
arise from the superposition of two populations in the dSph with
different velocity dispersions and spatial scale-lengths. While this
is certainly an interesting suggestion, we note that all our assumed functional
fits to the observed dispersion profiles are statistically consistent
with the observed data.

Given these caveats, we conclude that the Jeans analysis demonstrates
that the observed velocity dispersion profiles and cored light
distributions of dSphs are likely to be consistent with their
inhabiting dark matter haloes with central cores. We note that there
is no reason that the photometric scale length is exactly the
underlying mass scale length. From the mass modelling, the lower limit
on mass density core size is constrained to be at most a factor two
smaller than the observed luminosity core in UMi and
Fornax. Similarity, rather than exact equality, of the two scales is
what is relevant here.  As discussed below, however, cusped mass
distributions can also reproduce the observed data on the light
profile and velocity dispersion profile. To determine the actual
slopes of the inner dark matter density profiles, further information,
either in the form of larger velocity data sets which permit full
distribution function modelling (see below) or complementary dynamical
evidence is required. Fortunately, in two special cases, those of Ursa
Minor (UMi) and Fornax, there is additional information that enables
us to distinquish between shallow and steep internal mass density
profiles.

In UMi, an otherwise very simple system from an astrophysical
perspective, an extremely low velocity dispersion sub-structure
exists. \cite{kleyna03} explain this as a star cluster, which has
become gravitationally unbound (the normal eventual fate of every star
cluster), and which now survives as a memory in phase space. Why does
it survive in configuration space? The group of stars has the same
mean velocity as the systemic velocity of UMi, so it must orbit close
to the plane of the sky, and hence through the central regions of
UMi. As \cite{kleyna03} show, persistence of the cold structure is
possible only if the tidal forces from the UMi central mass gradient
are weak. In fact, survival of this phase-space structure in
configuration space requires that UMi has a slowly varying inner mass
profile, that is, a core, rather than a cusp.

The Fornax dSph galaxy has five surviving globular clusters. The orbit
of a compact massive system, such as a globular cluster, should decay
due to dynamical friction as it orbits through the background dark
matter halo particles. The rate of this orbital decay is faster in a
steep (cusped) dark-matter density profile, and is slower in a shallow
(core) dark-matter density profile. The (projected) distribution of
the surviving clusters has been analysed most recently by
\cite{Goerdt06}, who show that a cored mass distribution is strongly
preferred.

In summary, while Jeans' equation dynamical analyses cannot be
assumption independent, in both cases where some independent
information is available, shallow (cored) mass distributions are
preferred. There is no case where a steep (cusp) distribution is
required by the data. Applying Occam's razor, we therefore assume for
the remainder of this analysis that all dSph have similar underlying
dark matter mass profiles, which are cored. We adopt this specific
case-dependent result as a general result since it provides a natural
context for a characteristic length scale, which is suggested by
Figure~1. A single mass model, supported in two specific cases with
suitable data, links two otherwise disparate results, one
photometric, one kinematic.

\subsection{King-model dynamical analyses}

Dynamical analysis requires some simple and mathematically smooth
functional description of the spatial distribution of the stellar
tracers of the gravitational field. While any smooth function is
adequate, those most frequently used include Plummer models and King
models. Each is convenient, but can mislead if the parameters in the
adopted fitting function are (over)interpreted as having physical
meaning.

The King model (King 1966; see also Binney \& Tremaine 2000) is
physically valid for a self-gravitating system with a velocity
distribution function which is a lowered Maxwellian, i.e.~approximates
an isothermal distribution at small radii, with an imposed small core
- to avoid an unphysical divergence - and an imposed cutoff at large radii
 - to prevent infinite extension, and infinite velocities.  This
model is a good description of a stellar globular cluster, in which
mass follows light. The equilibrium velocity distribution function
which underlies this model is that generated by cumulative long-range
gravitational interactions between the component stars, which brings
a system of $N$ stars with rms velocity dispersion $v$, scale size
$R$, and corresponding crossing time $t_{cross}$ into dynamical
equilibrium in a characteristic time 
$\approx N/8ln(N) \times t_{cross}; t_{cross} \approx R/v$. It is
therefore naturally appropriate to any system in which the
astrophysical lifetime is long compared to the dynamical relaxation
time, and in which stars are orbiting in a gravitational potential
generated self-consistently from their mass.

We review the relevant range of applicabilities in Table~2. Only the
systems in that Table with dynamical age $\gta 1$ are amenable to a
physically meaningful King-model analysis, where we define `dynamical
age' as the ratio of the astrophysical age to the relaxation
time. dSph galaxies have a dynamical age 3-4 orders of magnitude
outside this range of validity, being too young in a dynamical sense.

We may ask what would be the stellar mass of a system with the
internal velocity dispersion and spatial scale size of a
representative dSph galaxy, and which has its dynamical relaxation
time less than the Hubble age of the Universe, so that the physical
conditions appropriate to establish a King model are in place. The
result is that one requires a stellar system of $5\times10^{10}$
stars. For a plausible stellar mass to light ratio, this implies a
galaxy of luminosity $\gta 10^{10}\Lsun$. Observed dSph galaxies are
many orders of magnitude less luminous. Thus, one does not expect that
a King model will be a physically valid description of a dSph galaxy,
although, as Table~2 shows, one does expect such models to be
reasonable approximations for massive star clusters. While it may
still be convenient to use a King model as a fitting function for a
galaxy, it is invalid to interpret the two free parameters of that fit
- the `core radius' and the `tidal radius' -- as meaningful physical
properties of the galaxy. \cite{Wu2007} also shows that King models
are inadequate descriptions of available dSph kinematics.

More generally, there are unavoidable consistency requirements in any
mass follows light model. In any model where mass follows light the
projected velocity dispersion must be maximum at the centre, and then
fall monotonically.  For a well-mixed (star cluster) system, the
velocity dispersion will decrease by roughly a factor of two over three
core radii.  This is an unavoidable requirement for any mass
follows light system, and is observed in star clusters (Figure~2).
Such a velocity dispersion profile is not required by data in any
well-studied galaxy, however small, further emphasising the intrinsic
difference between (virial, King) star clusters and galaxies.

\subsection{Time-dependent kinematics and radial range of valid analysis}

Determination of the mass distribution in the outer parts of a dSph
satellite galaxy remains a complex and data-starved challenge. The
(very few) tracer stars at large radii occupy the extreme limits of
the kinematic distribution function, where simplifying assumptions are
least reliable. At some radius tidal forces from the Milky Way must
become important, violating the equilibrium dynamics assumption,
though that radius depends on the {\sl a priori} unknown dSph dark matter
distribution and the (usually poorly constrained) orbit of the dSph
around the Galaxy. 

Very many studies are available predicting the effects of
time-dependant tides on the structure and kinematics observable in the
outer parts of galaxies. However, most such studies are idealised, and
consider only single-component models - i.e.~either no dark matter, or
only dark matter, as one prefers \citep{JSH99, JCG, Sohn2006}. This
makes any comparison to observations of the stellar density
distribution or of the stellar kinematics in a dSph galaxy with a dark
matter halo at best problematic. One recent example, among several,
which illustrates the richness of the potential tidal effects on stars
orbiting in dark-matter potentials is that of \cite{Read2006}, which
is one of the few studies to consider external tidal effects on
two-component dSph models in which stars orbit inside a dark-matter
halo. In another study of tides, \cite{Kl03} model the available data
for the Draco dSph, under the assumption that there is no dark matter
(ie a single-component model), so that the dSph is unbound.  They
deduce that the present smoothness and small line-of-sight depth of
this galaxy make unbound models impossible, and conclude that
dynamically dominant dark matter is required.

The observed kinematic and spatial distributions in the outer parts of
the dSph remain poorly determined by observations. As may be seen in
Figure~2, there is some evidence for low-dispersion (cold) outer
populations (\cite{W04}), and also for flat velocity dispersion
profiles to large distances (\cite{Sohn2006, Munoz2005}), both of
which are inconsistent with simple tidal disruption effects,
particularly since most dSphs show no evidence of apparent rotation
\cite{Koch06} , another prediction of tidal disruption models (Read et
al.~2006). Carina may be an exception, as \cite{Munoz06a} have
recently detected a velocity gradient on large scales (beyond the
nominal King-model fit `tidal' radius) in this dSph.  Photometric
studies continue to be inconclusive, with some suggesting a
characteristic signature of tidal distortions
\citep[e.g.][]{Ir95,Sohn2006}, but with later studies of the same
galaxy failing to find any signal from improved data (Odenkirchen et
al (2001), Segall et al (2006)).  Considerable uncertainty about the
dynamical state of the three dSphs nearest to the Galactic centre however
remains, with the closest one - Sgr - being manifestly disrupted.

Fortunately for our present purposes, which are concerned with the
properties of the mass distribution in the inner regions of the dSph,
the dynamical state of the very outer parts of the dSph is relatively
unimportant. The complexities noted do mean any discussion of total
masses is currently impracticable.

We also note for now that if indeed it is shown from future studies
that dSph galaxies are {\sl \bf not} strongly dark matter dominated,
but are star-cluster like stellar systems whose structural and
kinematic properties have been inflated in some way, the conclusions
of this paper concerning a minimum scale length on which dark matter
is seen to concentrate are inevitably and drastically strengthened. The
minimum dark matter clustering scale would have to extend beyond that
derived here, and become of order 1kpc.

\subsection{Distribution function modelling}

A known limitation of Jeans' equation moment analyses is that (at
least) some of the dSph galaxies show complex stellar populations, so
that adoption of a single dispersion profile and single length scale
is necessarily a simplification.  In all cases these Jeans' moment
analyses are applicable only over the range where simple functions
provide an adequate description of the underlying galaxy, which in
general is limited to one to two physical (luminous) scale
lengths. More complex behaviour, preventing valid application of such
simple models, is seen in the very outer parts of most dSph studied to
date, where also identification of member stars becomes increasingly
uncertain.  In general, a more robust analysis requires
significantly more information.

As larger data sets become available distribution function modelling
can supersede use of the Jeans' moment equations.  In distribution
function analyses one proceeds by constructing parameterised
equilibrium dynamical models, allowing the dark halo shape and mass,
and the tracer velocity anisotropy to vary. From these models one can
determine model distributions of the observable line of sight
velocity.  These models are then convolved with observational errors
and an orbital velocity distribution for binary star systems
appropriate to a dataset of interest, to predict observable velocity
distributions at every point across the projected galaxy.  It is then
straightforward to determine the best fitting models using the
individual stellar velocities, without the need to degrade the data
into moments (dispersions).

Models of this type have been applied by \cite{kleyna01} and by
\cite{Walker06}, and are being developed further for future
application to all large available data sets.  `Mass follows light' 
models for Draco were ruled out at the $2.5\sigma$ confidence level
using this type of analysis (\cite{kleyna01}).  Constant anisotropy
models of this type favour rather shallow halo inner mass profiles
with $\rho \propto r^{-0.5}$~\citep{Magorrian03, Koch06}.

A form of distribution function modelling has recently been used by
\cite{PMN07}, who apply a methodology developed by Lokas (Lokas 2002;
Lokas, Mamon \& Prada 2005) to fit observed dispersion profile data
adopting a King-model for the luminous galaxy, and embed this inside
an assumed NFW dark halo. This analysis requires, as is usual in such
fitting of NFW models, very considerable dark masses associated with
the observed dSph. NFW virial masses for the dwarfs considered in this
paper are all larger than $10^9\Msun$, with that for Draco having
log(mass)=9.8$\Msun$. This form of modelling breaks the degeneracy in
the mass determination discussed above essentially by requiring that
the assumed dark halo be of NFW form.

\citet{Wu2007} has recently applied more general distribution function
models to a rebinned version of the data for Draco and UMi shown in
our Figure~2 here (excluding our outermost data), and for other
published data for the Fornax dSph. Wu's analysis considers both two-
and three-integral distribution-function models, assuming the galaxies
are stable axisymmetric systems embedded in spherical dark-matter
potentials. He includes a range of forms for the underlying
gravitational potential, considering constant density models, constant
mass-to-light ratio models, dominant central point-mass (massive black
hole) models, NFW models, power-law and isochrone models. Wu's
isochrone model is closest in form to those derived here in our Jeans'
analysis (Figure~4), having an inner cored mass profile out to some
radius determined by fitting to the data, beyond which the profile of
the dark matter distribution steepens.

Wu's analysis shows that all of the NFW, power-law and isochrone
models are consistent with the data, while the constant density,
constant mass-to-light ratio and dominant central black hole models
are strongly ruled out. This conclusion is in very good agreement with
results of our analysis in this paper, where we use additional
information to prefer cored mass models (isochrone-like potentials)
over the NFW and pure power-law cases. Wu's characteristic scale
lengths at which the underlying dark matter density breaks below the
cored inner distribution are 500pc, 200pc and 900pc for Draco, UMi and
Fornax, respectively. These results are quite consistent with the
results of the simpler Jeans' analysis we present here.

\subsection{Mass distributions in dSph galaxies}

There are several Jeans' equation analyses of dSph kinematic data
which are fully described in the recent literature.  Figure~4
summarises the results of these Jeans equation models for several of
the dSph from \cite{W06a}, with more recent results for Leo~I from
Koch et al (2006) and for Leo~II from Koch et al
(2007), with in each case the simplest possible assumptions for the
velocity distribution, namely that it is isotropic at all radii.  It
is apparent that the models are invalid at large radii, where an
unphysical oscillation in some of the mass profiles is derived. In the
inner regions the models are well-behaved, and reproduce the overall
shape of the observed dispersion and light profiles. As illustrated in
detail in, for example, Koch et al (2006; especially their figure~11),
for Leo~I, both cored and cusped mass models can provide acceptable
agreement with the data for a suitable value of a constant anisotropy, and
excellent agreement when allowing a radially variable stellar orbital
anisotropy. As described above, although the Jeans models alone cannot
distinguish between cored and cusped mass distributions, in those two
cases where additional information is available cored density profiles
are preferred.

In Table~3 we summarise the mass determination results available. The
most robust numbers are the cumulative mass within the extent of the
kinematic data, and the associated mean density and outer circular
speed (columns 5, 6, 8 in Table~3). The central density is more
model-dependent. As an illustration of the range of likely values,
column 7 gives the mean density within 10pc in the case in which the
dark matter has a density profile which goes as $\rho \propto r^{-1}$
throughout the volume occupied by the stellar distribution. If the
dark matter has approximately constant density out to some break
radius, as our models prefer, then its central density will of
course, be comparable to the mean density quoted in column (6).  Table~3
also provides an estimate of the mean phase space density within the
half-light radius, which is defined as $3/(8\pi G r_{\rm h}^2
\sigma)$. This estimate assumes that the half-mass radius is
comparable to the observed half-light radius and that the stellar
velocity dispersion is related to the total mass through $\sigma^2 = G M /
r_{\rm h}$. Since neither assumption is strictly valid (columns (3)
and (5) show that this yields an underestimate of the true mass)
column (9) should be interpreted as an order of magnitude estimate only.

The total masses within the optical radii of the dSph galaxies have
been suspected for some years of showing a remarkably small range. 
Mateo (1998) showed that the available data at the time
was consistent with an apparent minimum dark halo mass, within the
optical galaxy, of order $10^7\Msun$. This relationship was extended
and developed by Cote et al (1999), who showed it also applied to
available data from M31 satellites. Later updates were provided by
\cite{W06a} and by \cite{LADM}. The current version of that
relationship is shown in Figure~5. Remarkably, the Mateo proposal has
survived an increase in the dynamic range of the sample by an order of
magnitude in both axes, and has become better-established, and of
lower scatter, as newer data have become available.

The Mateo plot presents total dark masses within the optical radii.  A
total mass is the integral over a scale length, a mass density
profile, and a central mass-density normalisation. Each of those
parameters is addressed in this paper. The photometric scale lengths
are summarised in Figure~1, which illustrates that dSph galaxies have
both a minimum scale size ($\sim120$pc), and a rather small range of
scale sizes. The available kinematic data and their analyses are
discussed above, summarised in Figure~4. That shows that each galaxy
studied, under the assumption of a valid Jeans'-equation analysis, has
a similar mass profile, both in shape and in normalisation: in at
least two cases it is probably cored, and the central density,
assuming a cored profile, is then very similar for all cases analysed
to date. For clarity, we note that have no robust independent proof
that the photometric scale length is exactly the underlying mass scale
length. From the mass modelling, the lower limit on mass density core
size is constrained to be at most a factor two smaller than the
observed luminosity core in UMi and Fornax. Similarity, rather than
exact equality, of the two scales is what we are assuming here.
Constancy of the scale length, the normalisation, and the density
profile parameters naturally explains the Mateo plot, which is their
product.

\subsection{ Are these derived properties robust?}

The photometric (length-scale) data are described above, and shown to
be robust for star clusters, and conservative relative to the present
conclusions for galaxies.  

There are two dynamical effects of relevance to bear in mind. In dense
star clusters the very steep and highly time-dependent gravitational
potential gradient between binary stars (especially close binaries)
provides a force which could eject dark matter particles. Essentially,
close binary stars will orbit through the cluster threshing the local
potential gradient,  clearing a dynamical tunnel through
phase-space, and ejecting any dark matter particles which might have
been present. Such an effect will be particularly important in the
central regions of star clusters, which are continually occupied by
close binaries. This effect is worth additional
consideration, and suggests that more extensive numerical modelling
and kinematic studies of the outer parts of diffuse star clusters is
worth while. Significant dynamical evolution will be irrelevant in
systems whose internal dynamical relaxation time is much longer than
its age. We show above that all the small galaxies of interest here in
fact will be immune to this effect.

The second important consideration concerns the validity of the
assumed steady-state dynamical analysis. There is continuing debate
and study of the possible relevance of time-dependence in dynamical
analyses of dSph galaxy stellar kinematics. Time-dependent tidal
disruption is manifestly dominant in the most nearby dSph galaxy, Sgr,
and is probably relevant for other dSph galaxies within a few tens of
kpc from the Galactic centre (ComaBer and UMaII are the obvious
candidates).  In the more distant galaxies there is no kinematic
evidence that the internal stellar kinematics in their central regions
are in any way affected by external Galactic tides
\citep{Koch06}. Most - but not all - kinematic studies remain limited
to moderate sample sizes, so that necessarily statistical and
model-dependent removal of interlopers and outliers can affect
conclusions somewhat (eg \cite{Klimentowski07}). 
Similar conclusions have very recently been derived by Wu
\citep{Wu2007} in his parametric re-analysis of published data for Draco,
Fornax and UMi. Wu concludes: `` Because [his distribution function]
models can fit both radial velocity profiles and surface number
density profiles, the so-called ``extra-tidal extensions'' in the
surface number density profiles found by \citet{Ir95} and \citet{W04}
do not require any special ad hoc explanation. Thus it is not valid to
consider them as the evidence of tidal stripping, as proposed by some
authors \citep[e.g.,][]{Mar01, Go03, Munoz2005}.''

We note for completeness that there are continuing efforts to apply
pure tidal models -- i.e. there is no dark matter, and the velocity
dispersions are inflated by tides -- to dSph kinematics.  If these
models can be proven to be relevant, the implications for the nature
of dark matter are quite profound. Substantially longer minimum length
scales and substantially lower maximum mass densities even than those
derived here will be required, providing quite extreme constraints on
the nature of dark matter. It remains far from clear that such
mass-follows-light models are consistent with the robust evidence on
cosmological scales that dark matter dominates the Universe.

\section{Discussion and Implications}

There is strong, and strengthening, observational evidence for four
conclusions concerning the smallest galaxies and typical or brighter
(globular, nuclear) star clusters. These two families of objects
co-exist over the range of stellar absolute magnitudes from $ -15 \lta
M_V \lta -4$, corresponding to luminosities from $\sim10^9\Lsun \gta L
\gta 10^3\Lsun$, with the upper limit corresponding to the brightest
known star clusters [\cite{SDHD06, Hilker07, EGDH07}], the lower
limit to the least luminous galaxies as yet discovered
(\cite{catsdogs}).  The number of well-studied objects known within
these limits in both classes has increased substantially in recent
years, providing an adequate sample to identify systematics, and to
test earlier, provisional trends (\cite{Cote99, W06a}).

We conclude the following:

{\bf ONE:} Over this substantial dynamic range, there is a clear
bimodality in the size distributions of the two families of object:
all smaller objects are star clusters, and have characteristic scale
sizes $\lta 30$pc ($\lta 10^{18}$m), while all larger objects are
galaxies, and have scale sizes for their luminous (stellar) components
$\gta 120$pc ($\gta 4.10^{18}$m).

{\bf TWO:} Where kinematic studies exist [essentially in the central
luminous regions], there is a clear distinction between the
phase-space distribution functions of star clusters and galaxies. At a
given (stellar, baryonic) luminosity, galaxies have phase-space
densities typically two to three orders of magnitude lower than do
star clusters.  In all adequately-studied cases, the galaxy's luminous
stellar component is embedded in a more extended dark matter halo. 
Star clusters over the whole mass range are well-described by
the virial theorem, and show no evidence for any (dynamically
significant, extended) dark matter halos.

{\bf THREE:} In the two specific galaxies where both detailed
dynamical analyses are feasible, ie, substantial kinematic data across
the face of the galaxies is available, and where independent evidence
to break the core/cusp/velocity-anisotropy mass degeneracy exists, the
derived dark mass distribution has a shallow density profile. For a
density distribution $\rho$ describable as a function of radius $r$ as
$\rho(r) \propto r^{-\alpha}$, the data imply that the power-law index
$\alpha \lta 0.5$, and with $\alpha$ being consistent with zero in the
innermost regions. Simplicity argues for this being the general
case. Under that assumption of generality, the minimum photometric
length scale can be interpreted as comparable to a minimum length
scale for the clustering of dark matter. The lower limit on the size
of a mass density core is constrained to be at most a factor two
smaller than the observed luminosity scale, while our dynamical
modelling suggests it is not much, if at all, larger.

{\bf FOUR:} Using the mass-model assumed in point THREE, the derived
mean mass density of dark matter within one to two half-light radii
for all the galaxies is $\rho_{\rm DM} \lta 5{\rm GeV/c}^2{\rm
cm}^{-3}$ (about 0.1$\Msun$ pc$^{-3}$). If we do not adopt the results
of point three, but allow a cusped mass model, the derived maximum
mass density of dark matter within 10pc of the centre is $\rho_{\rm
max,DM} \lta 2{\rm TeV/c}^2{\rm cm}^{-3}$ (about 60$\Msun$ pc$^{-3}$).

The combination of the small range of observed scale sizes, together
with an apparently standard (isochrone-like) form for the derived dark
mass density profile and its normalisation, naturally explains the
observed relation that all dSph galaxies have similar total dark mass
within their optical radii.

These systematic properties have obvious implications for the nature
of dark matter and galaxy formation in small halos, which we note very
briefly here. The simplest case is that our cored mass profiles are
the unmodified outcome of formation of small halos. An alternative is
that dark matter profiles on these small scales have been drastically
restructured by some astrophysical (feedback) process. We very briefly
note each in turn.

\subsection{ Implications for $\Lambda$CDM galaxy formation models}

One immediately asks if our observed length scale and characteristic
upper mass density derived here for the intrinsic nature of dark
matter is consistent with the extremely high-quality agreement between
$\Lambda$CDM models and large-scale structure, assuming the shallow
small-scale profiles are an intrinsic feature of dark matter halo
formation.Few (if any) cosmological observations of large-scale
structure have resolution on
small enough scales to be sensitive to the truncation of the
small-scale power spectrum on scales of order 100pc implied here, so
there is no disagreement.  Galaxy formation models inside the
$\Lambda$CDM paradigm however have considerable difficulties matching
observations on small scales. The well-known `satellite problem' is an
example, as is the `cores {\sl vs} cusps' debate. The amount of
structure on sub-kpc scales predicted by simulations will be very
drastically modified by our conclusions here, which imply that the
primordial power spectrum is truncated at small physical length
scales. In both the satellite counts and the core-cusp debates, the
discordance between data and numerical models will be greatly reduced
in simulations which include the minimum scale cutoff suggested by our
results. Numerical studies are underway to quantify this.

A fundamental astrophysical, rather than dark matter, puzzle which is
not simply explained is the now well-established bimodal size
distribution function illustrated in our Figure~1. Star clusters have
a maximum half-light scale size of some 30pc, while the characteristic
minimum scale length associated with the stellar systems which occupy
dark matter halos is some 4-5 times larger.  The dark matter mass
profiles derived here have a scale length always a factor of several
longer than the scale on which self-gravitating star clusters
form. But why do stellar systems not form with all possible scale
lengths inside the shallow dark matter potentials which we observe
kinematically? It is a generic prediction of $\Lambda$CDM galaxy
formation models that the baryonic component should cool and collapse
by a factor $\lambda^{-1} \approx 10$ more than does the (presumed
self-interaction-free) dark matter. Why is this not seen in these very
low-mass halos?

It seems that stellar populations in these shallow potentials expand
to a scale size comparable to that of the underlying potential. An
attempted explanation is beyond our present purposes. We note however
that there are many studies of feedback in shallow potential wells
(see eg. \cite{DS86,SWS87,RPV2006,Mayer2007}) which discuss the clear
importance of energy balance in driving gas loss and in expanding
extant stellar systems. We have attempted an explanation of this
ourselves (\cite{RG2005}), concluding that plausible astrophysical
feedback capable of leaving a dSph-like remnant cannot convert a
cusped DM profile into a cored DM profile, but extended shallow
(exponential) dSph luminosity can be readily formed.  More generally
\cite{RPV2006} provide an extensive discussion and introduction to the
literature. Several of these models remove all residual gas in an
early star formation episode, with the lost gas-mass reducing the
binding energy of the residual stellar system, allowing it to expand
into the dark matter potential.  A further general constraint on these
promising models is that many dSph galaxies have supported extended
star formation, and some show evidence for internal chemical element
self-enrichment, sometimes over most of the age of the
Universe. Derived star formation rates tend to be very low - of order
one star per $10^5$ years - so that single disruptive gas-evacuation
events are not part of solution space \citep[e.g.,][]{HGV-G00,
  CHG02}. Such disruptive events are of course consistent with the
apparent absence of stellar systems occupying dark halos of lower mass
than those we observe, should they be more prevalent on smaller mass
scales that we study here.

\subsection{Implications for the nature of dark matter}

An adequate discussion is beyond our intention in this paper, which is
to establish the observational evidence.
The local dark matter mass
density near the Sun was determined using distribution function
modelling to have the value  $\rho_{\rm DM} \sim 0.3{\rm GeV/c}^2{\rm
cm}^{-3}$ (about 0.01$\Msun$ pc$^{-3}$)\citep{KG89, KG91}.
Our maximum mass density and
minimum physical length scale derived here imply  particle number
densities, and self-interaction cross sections (if one assumes
self-interaction is the physical cause of the present-day length
scale) which are readily calculable for any specific particle class.
We note only the obvious conclusion that the very low maximum mass
density derived here is challenging for models of dark matter which
are dominated by massive (of order TeV) particles, such as those
predicted by some supersymmetric theories.  Such particles would be
required to have a spatial number density $\lta 1$ cm$^{-3}$ even if
the mass density profile is cusped, and orders of magnitude lower if
our cored profiles are appropriate. Much lower-mass ($\mu$eV)
particles, such as the axion, would have correspondingly higher number
densities. Rather interestingly, intermediate-mass (keV) sterile
neutrino particles have been discussed (see eg \cite{DW94, Abaz01,
Kus06, BM07a, BM07b})as relevant in just the spatial and density range
we have derived here.

\section{Acknowledgement}

AK and EKG acknowledge support by the Swiss
National Science Foundation through grant 200020-105260

{}

\clearpage

\begin{deluxetable}{lccc}
\tabletypesize{\scriptsize} 
\tablecaption{Observed 
Properties of the Established Milky Way dSph Satellites and
Candidates. Quoted half-light radii are derived from Plummer model
fits from the identified source, except for those cases identified as
core radii. Core radii, from King model fits, are lower limits on the
half-light radius.
\label{tbl:pars}}  \tablewidth{0pt} 
\tablehead{
\colhead{Object} &  \colhead{L$_{tot}$} &
\colhead{Galacto-centric} & \colhead{Half-light Radius} \\
\colhead{name} &  \colhead{L$_{\odot,V}$} &
\colhead{distance (kpc)} & \colhead{pc}}
\startdata
Sgr\tablenotemark{a} & $\gta2\times10^7$ & 24 & $\gta500$ \\
Fornax\tablenotemark{1} &  $1.5 \times 10^7$ & 140 & 400 (core)\\
Leo~I\tablenotemark{2} & $4.8\times10^6$ & 240 & 330 \\
Sculptor\tablenotemark{3} & $2.2\times10^6$ & 80 & 160 (core) \\
Leo~II\tablenotemark{4}  &  $7\times10^5$ & 230 & $185$ \\
Sextans\tablenotemark{5} & $5.0\times10^5$ & 85 & 630 \\
Carina\tablenotemark{6}  &  $4.3\times10^5$ & 100 & 290 \\
UrsaMinor\tablenotemark{7} &  $3.0\times10^5$ & 65 & 300 (core)\\
Draco\tablenotemark{8}   &  $2.6\times10^5$ & 80 & 230 \\
CVn~I\tablenotemark{9}  & $\gta1\times10^5$   & 220 & 550 \\
Hercules\tablenotemark{10} & $\gta2\times10^4$ & 140 & 310 \\
Bootes\tablenotemark{11}  &  $\gta2\times10^4$ & 60 & 230 \\
Leo~IV\tablenotemark{10}  &  $\gta1\times10^4$ & 160 & 150 \\
UMa~I\tablenotemark{12} & $\gta1\times10^4$ & 100 & 290 \\
CVn~II\tablenotemark{10} &  $\gta7\times10^3$ & 150 & 135 \\
UMa~II\tablenotemark{13\,}\tablenotemark{a} & $\gta3\times10^3$  & 30
& $\sim$125 \\ 
\hline \\
ComaBer\tablenotemark{10} & $\gta2\times10^3$ & 45 & 70 \\
\hline \\
Segue~I\tablenotemark{10}\tablenotemark{b} & $\sim10^3$ & 25 & 30  \\
Willman~I\tablenotemark{14\,}\tablenotemark{b} & $\sim10^3$ & 40 & 20 \\
\enddata 
{
\tablenotetext{a}{Sgr certainly and UMa~II probably, the objects
  closest to the Galactic centre, are associated with
extended tidal streams: \cite{Be06a, Fe06b}. The quoted
parameters in both cases are highly uncertain.}
\tablenotetext{b}{Nature uncertain: suspected to be a globular star cluster}
\tablenotetext{1}{\cite{Walker06}; (2) \cite{Koch06}; (3) \cite{Mateo98,
    west06}; (4) \cite{Coleman2007}; (5) \cite{Kleyna04}; (6) \cite{W06a}; 
 (7) \cite{Ir95}; (8) \cite{W04}; (9) \cite{Zu06a, ICILM06}; (10)
  \cite{catsdogs}; (11) \cite{Be06b, Munoz06b}; (12) \cite{Willman05,
    KWEG05}; (13) \cite{Zu06c, Grill06}; (14) \cite{Will05}  } 
%
%
%
}
\label{tab:objs}
\end{deluxetable}

\clearpage

\begin{deluxetable}{lcccccc}
\tabletypesize{\scriptsize} 
\tablecaption{Dynamical relaxation times for dSph and star cluster densities
 \label{tbl:dyntimes}} \tablewidth{0pt}
 \tablehead{
\colhead{Star} &  \colhead{Radial} &
\colhead{1-D velocity} & \colhead{Crossing} &
 \colhead{Relaxation}
& \colhead{dynamical} & \colhead{object }  \\
\colhead{numbers} &  \colhead{scale (pc)} &
\colhead{dispersion (km/s)} & \colhead{time (yr)} &
 \colhead{time (yr)}
& \colhead{ age} & \colhead{type}  } 
\startdata 
 100 & 2 & 0.5   & $4\times10^6$ & $10^7$ &  $\gta 1$ &  open
cluster \\
 $10^5$ &  4 & 10 & $4\times10^5$ &  $4.10^8$ &  $\gta10$ &
 median glob cluster \\
 $10^8$ & 10 & 30 & $3\times10^4$ &  $2.10^{10}$ &  $\sim1$ &
nuclear/large glob cluster \\
 $10^{12}$ & $10^4$ & 300 & $3\times10^7$ &  $10^{17}$ &  $10^{-7}$
 &gE gal \\
 $10^{9}$ & $10^3$ & 50 & $2\times10^7$ &  $10^{14}$ &  $10^{-4}$ & dE gal\\
 $10^{8}$ & $400$ & 10 & $4\times10^7$ &  $10^{13}$ &  $10^{-3}$ & dSph \\
\enddata
\end{deluxetable}

\clearpage

\begin{deluxetable}{lcccccccccc}
\tabletypesize{\scriptsize} 
\tablecaption{Observed velocity dispersions for Milky Way satellites,
  and derived masses and densities. The columns are: (1) Galaxy name; 
(2) Total velocity dispersion from extant data (as quoted in associated
reference); (3) Crude mass estimate based on total velocity
dispersion; (4) Radial extent of mass models - the radial extent
quoted is either the actual region within which the mass has been
calculated (as given in the associated reference) or, in cases where
this radius is unavailable, the nominal King limiting radius (denoted
$r_{\rm lim}$); (5) Mass within radius (4) based on modelling of
extended dispersion profile; (6) Mean density within the radius in
column (4); (7) Mean density within 10pc assuming mass is distributed as a
power-law with $\rho \propto r^{-1}$; (8) Circular speed at edge of
data = $\sqrt{G M_{\rm dark}/r}$; (9) Estimated mean phase space density within
half-light radius $= 3/(8\pi G r_{\rm h}^2 \sigma)$. See text for an
explanation. 
\label{tbl:pars2}}  \tablewidth{0pt} 
\tablehead{
\colhead{galaxy} &  \colhead{$\sigma$} & \colhead{$r_{\rm
    h}\sigma^2/G$} & \colhead{$r_{\rm max}$} &  \colhead{M$_{\rm DM}$}
& \colhead{$\overline{\rho_{\rm DM}}(r_{\rm max})$} &
\colhead{$\overline{\rho_{\rm DM,cusp}}(10{\rm pc})$} &
\colhead{$v_{\rm circ}(r_{\rm max})$}  & \colhead{Phase-Space Density}\\ 
\colhead{name} & \colhead{km\,s$^{-1}$} & \colhead{$\Msun$} &
\colhead{kpc} &  \colhead{$\Msun$} & \colhead{GeV/c$^2$ cm$^{-3}$} &
\colhead{GeV/c$^2$ cm$^{-3}$} & \colhead{km\,s$^{-1}$} &
\colhead{$\Msun {\rm kpc}^{-3} {\rm km\,s}^{-3}$}} 
\startdata
Fornax\tablenotemark{1} &  $11.1\pm 0.6$ & $1\times 10^7$ & 1.5 & $\sim
3\times10^8$ & $\sim 0.7$ & $1.1\times10^3$ & 29 & $ 2\times 10^4$ \\ 
Leo~I\tablenotemark{2} & $9.9\pm1.5$ & $8\times 10^6$ & $r_{\rm lim} \sim 0.9$ &
$3-8\times10^7$ &  $\sim 0.3-0.9$ & $310-830$ & 12-20 & $ 3\times 10^4$ \\ 
Sculptor\tablenotemark{3} &  $7-11$ & $2 \times 10^6$ & $r_{\rm lim} \sim 1.8$ &
$\gta10^7$ & $\sim 0.015$ & 27 & - & $ 2\times 10^5$ \\ 
Leo~II\tablenotemark{4}  &   $6.8\pm0.7$ & $2 \times 10^6$  & $r_{\rm lim} \sim 0.5$
& $3\times10^7$ & $\sim 1.8$ & $\sim 930$ & 16 & $ 1 \times 10^5$ \\ 
Sextans\tablenotemark{5} &  $8$ & $9 \times 10^6$ & 0.8 & $\sim3\times10^7$ &
$\sim0.5$ & 390 & 13 & $ 9\times 10^3$ \\ 
Carina\tablenotemark{6}  &  $7.5$ & $4 \times 10^6$ & $r_{\rm lim} \sim 0.8$ &
$\sim4\times10^7$ & $\sim 0.7$ & 520 & 15 & $ 4\times 10^4$ \\ 
UrsaMinor\tablenotemark{7} &   $12$ & $1 \times 10^7$ & 0.5 & $\gta6\times10^7$ &
$\sim 4.5$ & $2.1\times10^3$ & 23 & $ 3\times 10^4$ \\ 
Draco\tablenotemark{7}   &   $13$ & $9\times 10^6$ & 0.5 & $\gta6\times10^7$ &
$\sim 3$ & $1.7\times10^3$ & 22 & $ 4\times 10^4$ \\ 
Bootes\tablenotemark{8}  &   $6.6\pm2.3$ & $2 \times 10^6$ & - &
$\sim10^7$  & - & - & -  & $ 8\times 10^4$ \\ 
UMa~I\tablenotemark{9}   &   $9.3$ & $6\times 10^6$ & - & $\sim10^7$ &
- & -  & -  & $ 4\times 10^4$ \\ 
\enddata 
{\tabletypesize{\footnotesize}
\tablenotetext{1}{\cite{Walker06}; (2)\cite{Koch06}; (3)
  \cite{Mateo98, west06}; \\ (4) \cite{Coleman2007, Koch07}; (5)
  \cite{Kleyna04,W06a}; \\ (6) \cite{W06a}; (7) \cite{W04}; (8)
  \cite{Be06b, Munoz06b}; \\ (9) \cite{Willman05, KWEG05}.  }
%
%
}
\label{tab:objs2}
\end{deluxetable}

\clearpage

\begin{figure*}[ht]
\begin{center}
\includegraphics[height=11cm,angle=270]{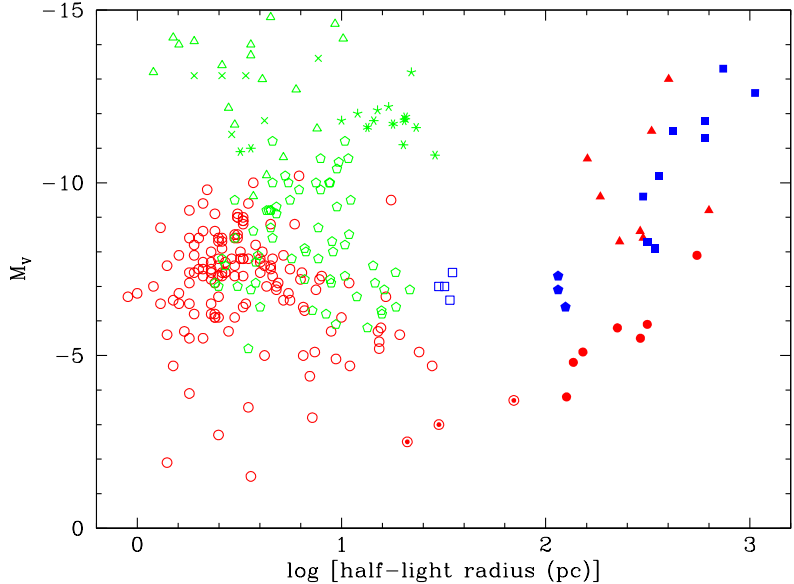}
\caption{Absolute magnitude ${\rm M_V}$ {vs} (logarithmic) half-light
  radius for well-studied stellar systems. The filled symbols are
  objects classed as galaxies, the open symbols and asterisks objects
  classed as star clusters of various types. Red colours indicate
  objects associated with the Milky Way Galaxy, blue colours are
  objects associated with M31 and green colours indicate more distant
  objects. Red filled triangles are the well-known dSph, red filled
  circles are those recently discovered, with in each case the
  photometry listed in Table~1 being adopted (references are given in
  the notes). The least-luminous M31 dSph (blue pentagons) are from
  Martin etal (2006), and have $\sim50\%$ uncertainties.  Ringed
  circles highlight the probable star clusters Segue~1 and Willman~1,
  and the object ComaBer. Open red circles are Milky Way globular
  clusters, from the compilation of Harris (1996), except the two
  largest Galactic globular clusters (Pal~5 and Pal~14) which use the
  most recent data from Hilker (2006). The largest globular clusters
  in M31 are shown as blue open squares, with data from Mackey et
  al.~(2006).  Green pentagons are globular clusters in NGC~5128 (the
  peculiar elliptical galaxy Cen A; Harris et al.~2002, 2006; Gomez et
  al.~2006). Green crosses represent nuclear star clusters in a range
  of external galaxies (Bastian et al.~2006), open green triangles are
  young massive star clusters (Bastian et al.~2006; Seth et
  al.~2006). Asterisks are Ultra Compact Dwarfs (UCD) in the Fornax
  cluster of galaxies (De Propris et al.~2005; Mieske et al.~2002;
  Drinkwater et al.~2003), and in the Virgo cluster (Hasegan et
  al.~2005). For UCD3* in Fornax we adopt the most recent core
  measurement (22pc) by Drinkwater et al.~2003).  Not shown
  individually are the ``Faint Fluffy'' star clusters found in the
  disks of lenticular galaxies (Brodie \& Larsen 2002) which have
  absolute magnitudes M$_V\sim -7$, and sizes in the range ten to
  twenty pc (1.0-1.3 in log(r$_h$)). Sgr is not shown. Half-light size
  definitions and determinations are discussed further in the text.  }
\end{center}
\end{figure*}

\clearpage

\begin{figure*}[!ht]
\begin{center}
\includegraphics[height=12truecm]{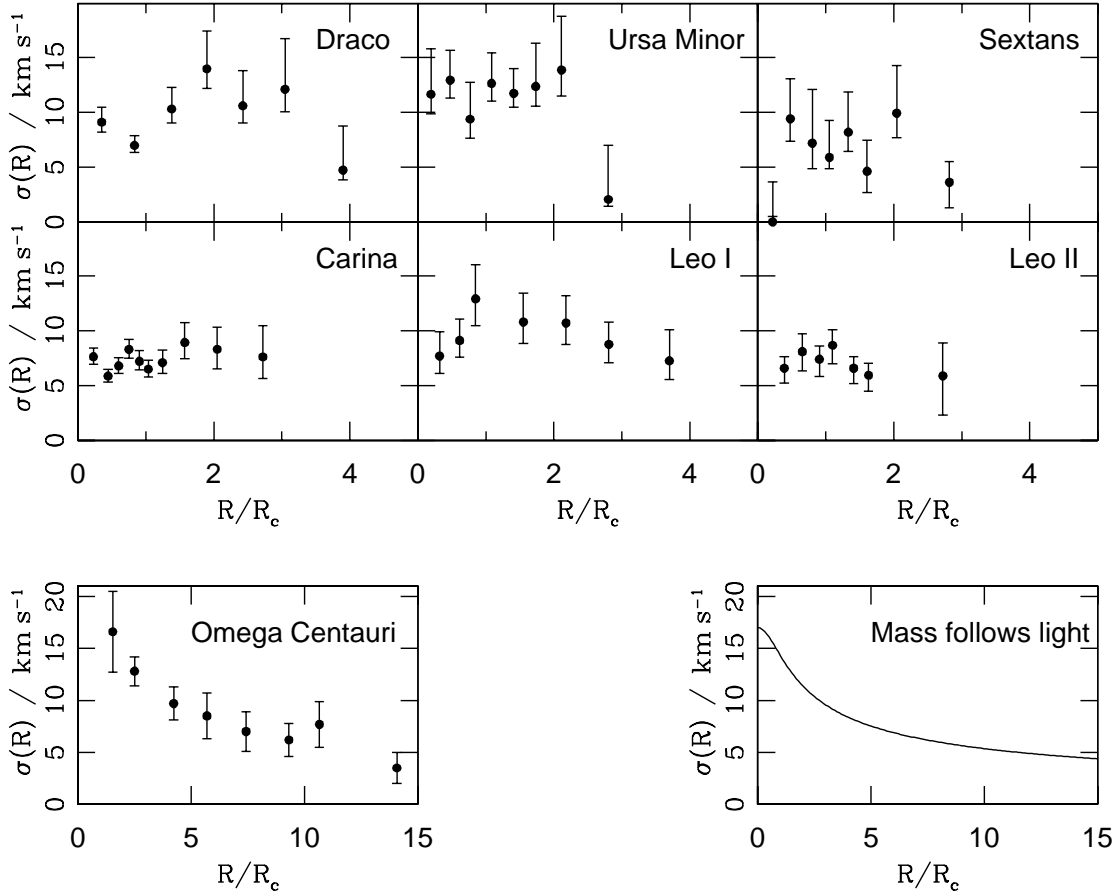}
\caption{Observed line-of-sight velocity dispersion profiles for six
dSph galaxies. Also shown (lower right) is the model predicted
dispersion profile for a Plummer model in which mass follows
light. The lower left panel shows the observed velocity dispersion
profile for the globular cluster Omega Cen
from \protect\cite{Seitzer83}.  The similarity between the Plummer
`mass follows light' model and the data for Omega Cen is apparent,
with a monotonic decrease in dispersion from a central maximum. In
contrast, the dSph galaxies do not have their maximum dispersion value
at the centre, and retain relatively high dispersions at large radii,
indicating extended (dark) mass distributions.  }
\end{center}
\end{figure*}

\clearpage

\begin{figure*}[!ht]
\begin{center}
\includegraphics[height=8truecm]{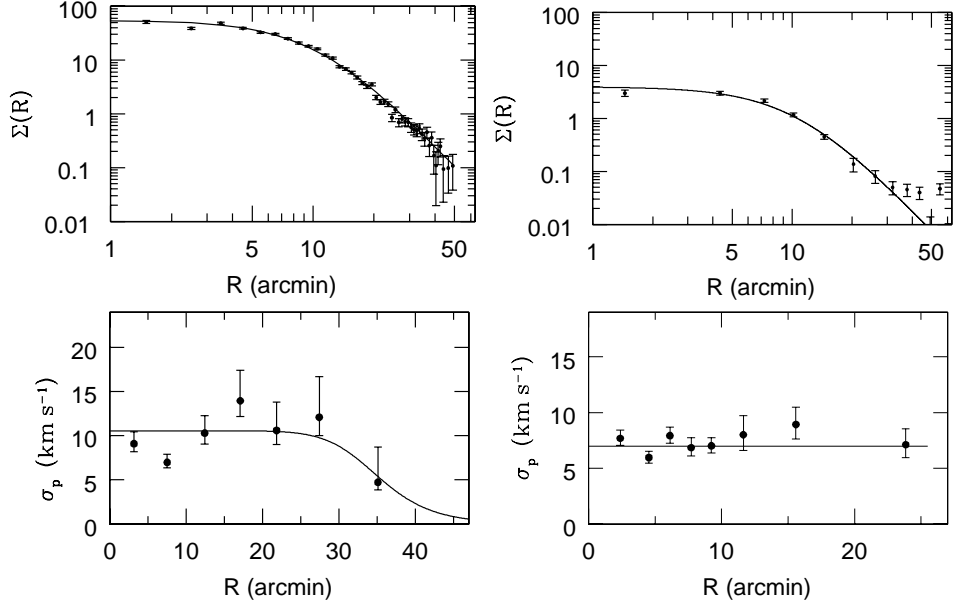}
\caption{Functional fits to the surface brightness profile (top) and
velocity dispersion profile (bottom) of the Draco (left panels) and
Carina (right panels) dSphs used to derive mass profiles based on
Jeans equations. Similar fits are used for the remaining four dSphs
presented in Figure~4.}
\label{fig:JeansFits}
\end{center}
\end{figure*}

\clearpage

\begin{figure*}[!ht]
\begin{center}
\includegraphics[height=10truecm]{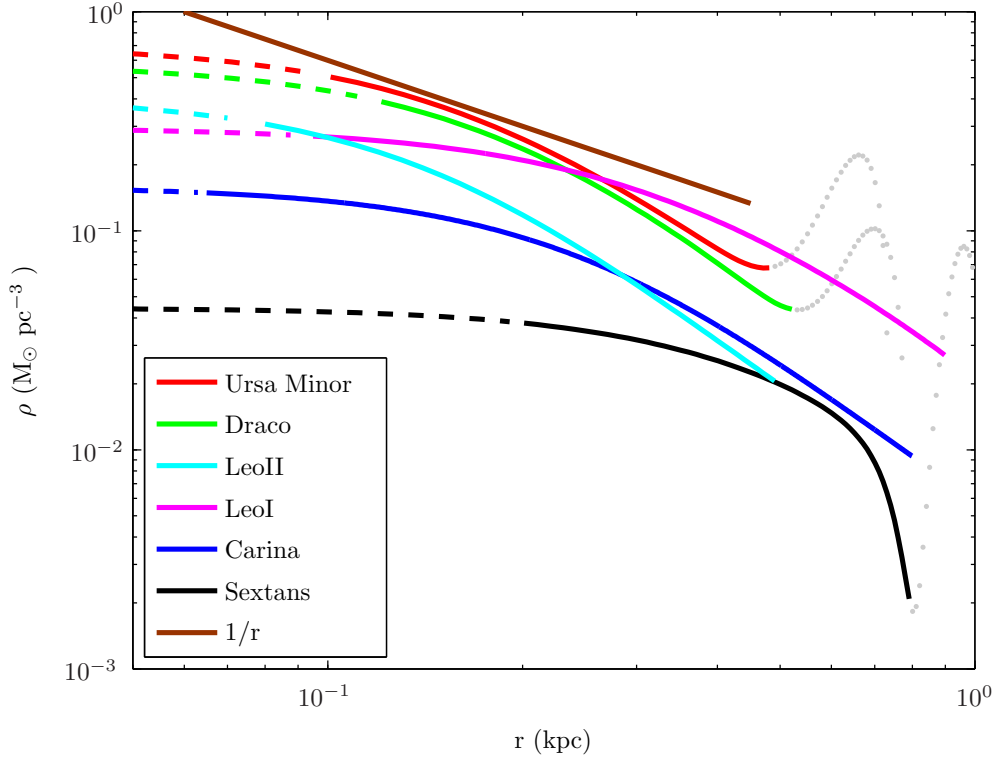}
\caption{Derived inner mass distributions from isotropic Jeans'
  equation analyses for six dSph galaxies. The modelling is reliable
  in each case out to radii of log (r)kpc$\sim0.5$. The unphysical
  behaviour at larger radii is explained in the text. The general
  similarity of the inner mass profiles is striking, as is their
  shallow profile, and their similar central mass densities. Also
  shown is an $r^{-1}$ density profile, predicted by many CDM
  numerical simulations (eg Navarro, Frenk \& White 1997). The
  individual dynamical analyses are described in full as follows: Ursa
  Minor (\cite{W04}); Draco (\cite{W04}); LeoII (\cite{Koch07}); LeoI
  (\cite{Koch06}); Carina (\cite{W06a}, and Wilkinson et al in
  preparation); Sextans (\cite{Kleyna04}).  }
\end{center}
\end{figure*}

\clearpage

\begin{figure*}[!ht]
\begin{center}
\includegraphics[height=10truecm]{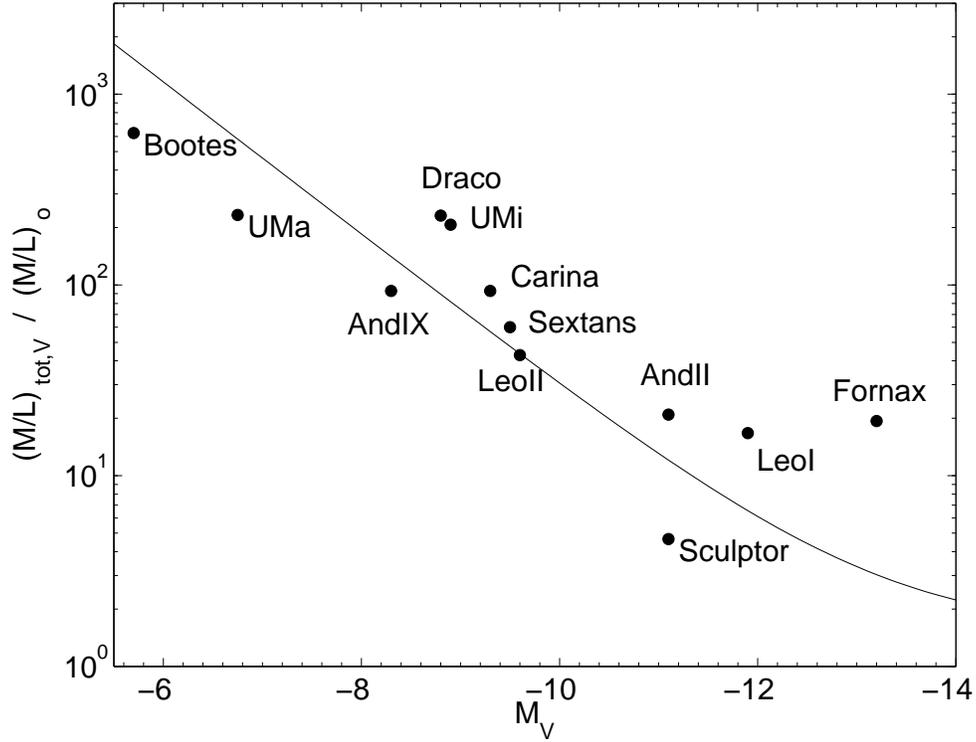}
\caption{An updated Mateo plot. Mass-to-light ratios are plotted
  versus absolute magnitudes for Local Group dwarf galaxies, following
  a style suggested by Mateo [Mateo etal (1993);(1998, his fig~9,
  lower panel)]. The solid line is the relation for a constant mass
  (dark) halo.  The modern data shown here extend the original
  relation by three magnitudes in luminosity, and an order of
  magnitude in mass-to-light ratio, while reducing the scatter by an
  order of magnitude. Data are from the tables in the text. Values for
  Scl, AndII, AndIX, UMa and Boo are based on small kinematic samples,
  and are less certain than are the results for the other galaxies.
  We explain this correlation as a consequence of the characteristic
  minimum galaxy scale size shown in Figure~1 convolved with the
  narrow range of mass profiles and mean dark matter densities shown
  in Figure~4.}
\end{center}
\end{figure*}


\begin{thebibliography}{}

\bibitem[Aaronson (1983)]{Aa83} Aaronson, M.,  1983, \apjl 266 L11

 \bibitem[Abazajian, Fuller, \& Patel (2001)]{Abaz01}Abazajian, K.,
Fuller, G.M., \& Patel, M., 2001 Phys Rev D {\bf 64} 023501

\bibitem[An \& Evans(2006)]{AnEvans06} An J.H., Evans N.W., 2006, ApJ, 642, 752 

\bibitem[Battaglia et al (2006)]{Batt06} Battaglia, G., etal 2006 \aap,
459 423

\bibitem[Belokurov et al.(2006a)]{Be06a} Belokurov, V. et
al.\ 2006a, \apjl, 642, L137 

\bibitem[Belokurov et al.(2006b)]{Be06b} Belokurov, V. et
al. 2006b, \apjl, 647, L111

\bibitem[Belokurov et al.(2007)]{catsdogs} Belokurov, V. et al 2007
  \apj, 654, 897

\bibitem[Bender etal (1992)]{bbf92} Bender, R.,  Burstein, D., Faber,
S., 1992 \apj,  399 462 

\bibitem[Bertin, G. (2000)]{Bertin} Bertin, G., Dynamics of Galaxies (CUP) 2000 

\bibitem[\protect\citeauthoryear{Biermann \& Munyaneza}
{2007a}]{BM07a} Biermann, P.L., \& Munyaneza, F., 2007a
astroph/0702164

\bibitem[\protect\citeauthoryear{Biermann \& Munyaneza}
{2007b}]{BM07b} Biermann, P.L., \& Munyaneza, F., 2007b
astroph/0702173


\bibitem[Binney, Bissantz \& Gerhard (2000)]{BBG} 
Binney,J.,  Bissantz \& Gerhard, O. 2000 \apj, 537 L99

\bibitem[Binney \& Tremaine(1987)]{BT87}
Binney J., Tremaine S. 1987, Galactic Dynamics, Princeton University
Press, Princeton

\bibitem[Brodie \& Larsen (2002)]{BL02} Brodie, J.P. \& Larsen,
  S. 2002, AJ, 124, 1410 

\bibitem[Brodie, Burkert \& Larsen (2006)]{BBL2006} Brodie, J. P., Burkert,
A., \&  Larsen, S. 2004 ASPC 322 139

\bibitem[Burstein, Bender, Faber \& Nolthenius (1997)]{bbfn} Burstein,D.,
Bender, R., Faber, S.,  \& Nolthenius, R. 1997 \aj, 114, 1365

\bibitem[Carigi, Hernandez \& Gilmore (2002)]{CHG02} Carigi, L.,
  Hernandez, X., \& Gilmore, G., 2002 \mnras 334 117

\bibitem[Cohen (2006)]{Cohen2006} Cohen, J.G. 2006  \apj, 653, L21

\bibitem[Coleman, Jordi, Rix, Grebel \& Koch (2007)]{Coleman2007}
  Coleman, M.G., Jordi, K., Rix, H.-W., Grebel, E.K., \& Koch, A. 2007
  AJ submitted

\bibitem[C\^ot\'e, Mateo, Olszewski \& Cook (1999)]{Cote99} C\^ot\'e, P., Mateo,
M., Olszewski, E., \& Cook, K., 1999 \apj, 526 147

\bibitem[\protect\citeauthoryear{Dekel \& Silk}{1986}]{DS86} Dekel,
A., \& Silk, J., 1986, \apj, 303, 39

\bibitem[De Propris et al.(2005)]{DeP05} De Propris, R., 
Phillipps, S., Drinkwater, M.~J., Gregg, M.~D., Jones, J.~B., Evstigneeva, 
E., \& Bekki, K.\ 2005, \apjl, 623, L105 

\bibitem[Dodelson \& Widrow (1994)]{DW94}Dodelson, S., \& Widrow, L.,
(1994) Phys Rev Lett {\bf 72} 17 

\bibitem[Drinkwater et al.(2003)]{Dr03} Drinkwater, 
M.~J., Gregg, M.~D., Hilker, M., Bekki, K., Couch, W.~J., Ferguson, 
H.~C., Jones, J.~B., \& Phillipps, S.\ 2003, \nat, 423, 519

\bibitem[\protect\citeauthoryear{Evstigneeva, Gregg, Drinkwater, \&
Hilker}{2007}]{EGDH07}Evstigneeva, E.A., Gregg, M.D., Drinkwater,
M.J., \& Hilker, M., 2007 astroph/0612483


\bibitem[Fall \& Rees (1977)]{FR77} Fall, S.M. \& Rees, M.J. 1977, MNRAS, 181, 37

\bibitem[Fellhauer et al.(2007)]{Fe06b} Fellhauer, M. et al. 2007
\mnras, 375, 1171

\bibitem[Gentile et al. (2004)]{Gentile2004} Gentile etal 2004 \mnras,
351 903 

\bibitem[Gentile et al. (2005)]{Gentile2005} Gentile etal 2005 
\apjl, 634 L145.


\bibitem[Goerdt et al. (2006)]{Goerdt06} Goerdt, T., Moore, B.,
Read, J. I., Stadel, J., Zemp, M. 2006 \mnras, 368 1073

\bibitem[Gerhard (2006)]{G06} Gerhard, O., 2006 astroph-0608343

\bibitem[Gilmore et al. (2006)]{LADM} Gilmore, G., Wilkinson, M., Kleyna, J.,
Koch, A.,  Evans, Wyn, Wyse,  R.F.G. \& Grebel,  E.K. 2006 astroph-0608528 

\bibitem[Gnedin \& Ostriker (1997)]{GO97} Gnedin, O. \& Ostriker,
  J.P. 1997, ApJ, 474, 223 

\bibitem[G{\'o}mez et al.(2006)]{Go06} G{\'o}mez, M., Geisler, D.,
Harris, W.~E., Richtler, T., Harris, G.~L.~H., \& Woodley, K.~A.\
2006, \aap, 447, 877

\bibitem[\protect\citeauthoryear{G{\'{o}}mez-Flechoso \&
Mart\'{i}nez-Delgado}{2003}]{Go03}G{\'{o}}mez-Flechoso, M. {\'{A}}, \&
Mart\'{i}nez-Delgado, D., 2003, \apj, 586, L123 

\bibitem[Green, Hofman \& Schwarz (2005)]{GHS05} Green, A.A., Hofman,
S., \& Schwarz, D.J., 2005 AIPC 805 431

\bibitem[Grillmair (2006)]{Grill06} Grillmair, C.J. 2006 \apjl, 645 L37

\bibitem[Harbeck et al.(2001)]{harbeck01} Harbeck, D., et al.\ 
2001, AJ, 122, 3092 

\bibitem[Harris (1996)]{Ha96} Harris, W.~E.\ 1996, \aj, 112, 
1487 

\bibitem[Harris et al. (2002)]{Ha02} Harris, W.~E., Harris, G.~L.~H.,
Holland, S.~T., \& McLaughlin, D.~E.\ 2002, \aj, 124, 1435

\bibitem[Harris et al. (2006)]{Ha06} Harris, W.~E., Harris, G.~L.~H.,
Barmby, P., McLaughlin, D.~E. \& Forbes, D.A.\ 2006, \aj, 132, 2187

\bibitem[Ha{\c s}egan et al. (2005)]{Ha05} Ha{\c s}egan, M., 
et al.\ 2005, \apj, 627, 203 

\bibitem[Hernandez, Gilmore \& Valls-Gabaud (2000)]{HGV-G00}
  Hernandez, X., Gilmore, G., Valls-Gabaud, D., 2000 \mnras 317 831

\bibitem[Hilker (2006)]{Hi2006} Hilker, M., 2006 A\&A 448 171 

\bibitem[\protect\citeauthoryear{Hilker etal }{2007}]{Hilker07} Hilker,
M., Baumgardt, H., Infante, L., Drinkwater, M., Evstigneeva, E., \&
Gregg, M., 2007 A\& A 463 119

\bibitem[Huxor et al. (2005)]{Hu05} Huxor, A.~P., Tanvir, 
N.~R., Irwin, M.~J., Ibata, R., Collett, J.~L., Ferguson, A.~M.~N., 
Bridges, T., \& Lewis, G.~F.\ 2005, \mnras, 360, 1007 

\bibitem[Ibata et al. (2006)]{ICILM06} Ibata etal 2006 
\mnras, 373 L70

\bibitem[Irwin \& Hatzidimitriou (1995)]{Ir95} Irwin, M., \&
Hatzidimitriou, D.\ 1995, \mnras, 277, 1354

\bibitem[Johnston, Sigurdsson \& Hernquist (1999)]{JSH99} 
Johnston, K., Sigurdsson, S., \& Hernquist, L. 1999 \mnras, 302 771

\bibitem[Johnston, Choi  \& Guharthakurta (2002)]{JCG}
Johnston, K., Choi, P.I., \& Guharthakurta, P., 2002  \aj, 124 127

\bibitem[King (1966)]{King1966} King I., 1966, AJ, 71, 64

\bibitem[Klessen et al. (2003)]{Kl03} Klessen, R.~S., Grebel, E.~K.,
Harbeck, D.\ 2003, \apj, 589, 798

\bibitem[Kleyna et~al. (2001)]{kleyna01}
Kleyna J.T., Wilkinson M.I., Evans N.W., Gilmore G. 2001, 
ApJ, 563, L115

\bibitem[Kleyna et al. (2002)]{kleyna02} Kleyna, J., Wilkinson, 
M.~I., Evans, N.~W., Gilmore, G., \& Frayn, C.\ 2002, \mnras, 330, 792 

\bibitem[Kleyna et~al. (2003)]{kleyna03}
Kleyna J.T., Wilkinson M.I., Gilmore G., Evans N.W.  2003, 
ApJ, 588, L21

\bibitem[Kleyna et al.(2004)]{Kleyna04} Kleyna J.T., Wilkinson M.I.,
  Evans N.W., Gilmore G., 2004, MNRAS, 354, L66  

\bibitem[Kleyna et al. (2005)]{KWEG05} Kleyna, J.~T., Wilkinson, M.~I.,
Evans, N.~W., \& Gilmore, G.\ 2005, \apjl, 630, L141

\bibitem[ Klimentowski et al. (2007)]{Klimentowski07} Klimentowski J.,
  Lokas E.L.,  Kazantzidis S., Prada F., Mayer L.,  
Mamon G.A., 2007, MNRAS, submitted, astro-ph/0611296

\bibitem[Koch et al. (2006)]{Koch06} Koch, A., et al 2007 \apj, 657, 241


\bibitem[Koch et al. (2007)]{Koch07} Koch, A., etal 2007 \aj, 133, 270

\bibitem[Kuijken \& Gilmore (1989)]{KG89} Kuijken, K., \& Gilmore,
G. 1989 \mnras, 239 605

\bibitem[Kuijken \& Gilmore (1991)]{KG91} Kuijken, K., \& Gilmore,
G. 1991 \apjl, 367 L9

\bibitem[Kusenko (2006)]{Kus06} Kusenko, A. 2006 hep-ph/0609158

\bibitem[Lin \& Faber (1983)]{LF83}  Lin,D., \& Faber,S.,  1983 \apjl,
266, L21

\bibitem[Lokas (2002)]{Lokas2002} Lokas, E.L., 2002 \mnras, 333, 697

\bibitem[Lokas, Mamon \& Prada (2005)]{LMP2005}
Lokas, E.L., Mamon, G, A., Prada, F. 2005 MNRAS 363 918

\bibitem[Mackey, et al (2006)]{dougalM31}  Mackey, A.~D., Huxor, A.,
  Ferguson, A.M, et al 2006 \apjl, 653 L105

\bibitem[Magorrian(2003)]{Magorrian03} Magorrian J., 2003, in
  R. Bender \& A. Renzini eds., The  
Mass of Galaxies at Low and High Redshift, ESO Astrophysics Symposia, 18

\bibitem[Martin et al. (2006)]{Ma06} Martin, N., et al. 2006, \mnras,
  371, 1983

\bibitem[\protect\citeauthoryear{Mart\'{i}nez-Delgado et
al.}{2001}]{Mar01}Mart\'{i}nez-Delgado, D., Alonso-Garc\'{i}a, J.,
Aparicio, A., \& G\'{o}mez-Flechoso, M. A., 2001, \apj, 549, L63 

\bibitem[Mateo, Olszewski, Pryor, Welch, \& Fischer (1993)]{Mateoetal1993}
Mateo, M., Olszewski, E.W., Pryor, C., Welch, D.L., Fischer, P. 1993
\aj, 105 510

\bibitem[Mateo (1997)]{mateo97} Mateo, M., 1997, in The Nature of
Elliptical Galaxies; 2nd Stromlo Symposium, ASP Conf.~Ser. 116, 259 

\bibitem[Mateo (1998)]{Mateo98} Mateo, M.~L.\ 1998, ARA\&A, 36, 
435 

\bibitem[\protect\citeauthoryear{Mayer, Katzantzidis, Mastropietro, \&
Wadsley}{2007}]{Mayer2007} Mayer, L., Katzantzidis,S.,  Mastropietro,
C., \& Wadsley, J., 2007 Nature 445 738

\bibitem[McConnachie \& Irwin (2006)]{Mc06} McConnachie, A.~W., \&
Irwin, M.~J.\ 2006, \mnras, 365, 1263

\bibitem[McConnachie, Pe{\~n}arrubia, \& Navarro
  (2007)]{McConnachie07} McConnachie A.W., Pe{\~n}arrubia J., Navarro
  J.F., ApJL, submitted, astro-ph/0608687

\bibitem[Meylan et al. (2001)]{Meylanetal2001} Meylan, G., etal \aj, 122 30

\bibitem[Mieske et al. (2002)]{Mi02} Mieske, S., Hilker, M., \&
Infante, L.\ 2002, \aap, 383, 823

\bibitem[Mu{\~n}oz et al. (2005)]{Munoz2005}
Mu{\~n}oz, R., Frinchaboy, P.M., Majewski, S.R.; etal 2005
ApJ  631 L137

\bibitem[Mu{\~n}oz et al.(2006a)]{Munoz06a} Mu{\~n}oz, R.R., et al.,
  2006a, ApJ, 649, 201 

\bibitem[Mu{\~n}oz et al. (2006b)]{Munoz06b} Mu{\~n}oz, R.~R., Carlin, 
J.~L., Frinchaboy, P.~M., Nidever, D.~L., Majewski, S.~R., \&
Patterson, R.~J.\ 2006b, \apjl, 650 L51

\bibitem[Navarro etal (1997)]{nfw} Navarro, J., Frenk, C. \& White,
  S.D.M. 1997, ApJ, 490, 493 

\bibitem[Odenkirchen et al. (2001)]{Od2001}
Odenkirchen, M., Grebel, E. K.. Harbeck, D., etal 2001 AJ
122 2538 

\bibitem[Odenkirchen et al. (2002)]{Od2002}
Odenkirchen, M., etal 2002 AJ 124 1497 

\bibitem[Odenkirchen et al. (2003)]{Od2003}
Odenkirchen, M., etal 2003 AJ 126 2385 

\bibitem[Ostriker \& Peebles (1973)]{OP73} Ostriker, J.P., \& Peebles,
J.  1973 \apj, 186  467

\bibitem[\protect\citeauthoryear{Ostriker \&
Steinhardt}{2003}]{jpops03} Ostriker, J.P., \& Steinhardt, P. 2003
Science 300 1909

\bibitem[\protect\citeauthoryear{Penarrubia, McConnachie \&
Navarro}{2007}]{PMN07} Penarrubia,J., McConnachie, A., \& Navarro, J.,
2007 astroph/0701780

\bibitem[Profumo, Sigurdson \& Kamionkowski (2006)]{PSK06} Profumo,
S., Sigurdson, K., \& Kamionowski, M., 2006 PHRvL 97 c1301

\bibitem[Pryor \& Meylan (1993)]{PM93} Pryor, T. \& Meylan, G. 1993 in
Structure and Evolution of Globular Clusters, ASP Conf Ser 50 357

\bibitem[\protect\citeauthoryear{Read \& Gilmore}{2005}]{RG2005}
Read, J.I.R \& Gilmore, G., 2005, MNRAS, 356, 107

\bibitem[\protect\citeauthoryear{Read, Pontzen \& Viel}{2006}]{RPV2006}
Read, J.I.R., Pontzen, A.P., \& Viel, M., 2006 MNRAS 371 885

\bibitem[Read et al. (2006)]{Read2006} Read, J. I., Wilkinson, M. I.,
Evans, N. Wyn, Gilmore, G., Kleyna, Jan T.  2006 \mnras, 367 387

\bibitem[Segall et al (2006)]{Segall06} Segall, M., et al 2006
  astroph-0612263 

\bibitem[Seitzer(1983)]{Seitzer83} Seitzer, P.O., 1983, Ph.D.~Thesis, 

\bibitem[Seth, Dalcanton, Hodge \& Debattista (2006)]{SDHD06} 
Seth, A.C., Dalcanton, J.J., Hodge, P. W., Debattista,
V.P., \aj, 132, 2539

\bibitem[\protect\citeauthoryear{Silk, Wyse, \& Shields}{1987}]{SWS87}
Silk, J., Wyse, R.F.G., \& Shields, G., 1987, \apj, 322, L59

\bibitem[Sohn et al. (2006)]{Sohn2006} Sohn, S.T., etal 2006 astroph-0608151

\bibitem[Tolstoy et al (2004)]{tolstoy2004} Tolstoy, E., et al 2004
  \apj, 617 L119

\bibitem[Unavane, Wyse, \& Gilmore (1996)]{Unavane} Unavane, M., Wyse,
R.F.G., Gilmore, G., 1996 \mnras, 278 727

\bibitem[Valenzuela et al. (2005)]{Valenzuela2005} Valenzuela etal astroph 0509644,

\bibitem[van de Ven et al. (2006)]{Ven2006} van de Ven etal 2006 \aap, 445 513 

\bibitem[Venn et al. (2004)]{Venn2004} Venn, K.~A., Irwin, M., 
Shetrone, M.~D., Tout, C.~A., Hill, V., \& Tolstoy, E.\ 2004, \aj, 128, 
1177 

\bibitem[Walcher et al. (2005)]{Walcher2005} Walcher, C.J., etal 2005
\apj, 618 237 

\bibitem[Walker et al. (2006)]{Walker06} Walker, M.G., etal 2006 \aj, 131 2114 

\bibitem[Westfall et al. (2006)]{west06} Westfall, K. B., etal 2006 \aj,
131  375

\bibitem[Wilkinson et~al.(2002)]{wilkinson02} 
Wilkinson M.I., Kleyna, J.T., Evans, N.W., Gilmore G. 2001, 
MNRAS, 330, 778

\bibitem[Wilkinson et~al. (2004)]{W04}
Wilkinson M.I.,Kleyna J.T., Evans N.W., Gilmore G., Irwin, M., Grebel,
E.K., 2004, ApJ, 611, L21

\bibitem[Wilkinson et al. (2006)]{W06a} Wilkinson, M.~I., Kleyna,
J.~T., Wyn Evans, N., Gilmore, G.~F., Read, J.~I., Koch, A., Grebel,
E.~K., \& Irwin, M.~J.\ 2006, EAS Publications Series, 20, 105

\bibitem[Willman et al. (2005a)]{Willman05} Willman, B., et al.\ 
2005, \apjl, 626, L85 

\bibitem[Willman et al. (2005b)]{Will05} Willman, B., et al.\ 
2005, \aj, 129, 2692 

\bibitem[Willman et al. (2006)]{Wi06} Willman, B., et al.\ 
2006, \aj, in press (astro-ph/0603486)

\bibitem[\protect\citeauthoryear{Wu}{2007}]{Wu2007}Wu, X., 2007
submitted to ApJ, astroph/0702233

\bibitem[Wyse, Gilmore, Houdashelt et al. (2002)]{WyseNA}
Wyse, R.F.G., Gilmore, G., Houdashelt, M.L., Feltzing,
S., Hebb, L., Gallagher, J.S., Smecker-Hane, T.A. 2002 NewAstron 7 395

\bibitem[Zaritsky, Gonzalez \& Zabludoff (2006a)]{ZGZ2006a} 
Zaritsky, D., Gonzalez.,  and Zabludoff, A. 2006 \apj, 638 725 

\bibitem[Zaritsky, Gonzalez \& Zabludoff (2006b)]{ZGZ2006b} 
Zaritsky, D., Gonzalez.,  and Zabludoff, A. 2006 \apjl, 642 L37

\bibitem[Zucker et al. (2004)]{Zu04}
        Zucker, D.~B. et al.\, 2004, \apjl, 612, L121
 
\bibitem[Zucker et al. (2006a)]{Zu06a} Zucker, D.~B., et al.\ 2006a, 
        \apjl, 643, L103

\bibitem[Zucker et al. (2006b)]{Zu06b} Zucker, D.~B., et al.\ 2006b,
        \apjl, submitted (astro-ph/0601599)

\bibitem[Zucker et al. (2006c)]{Zu06c} Zucker, D.~B., et al.\ 2006c,
        \apjl, 650, L41


\end{thebibliography}
\end{document}